\documentclass[final,times,number]{elsarticle}
\usepackage{physics}
\usepackage{amsmath}
\usepackage{comment}
\usepackage{float}
\usepackage{units}
\usepackage{amssymb}
\usepackage{makecell,diagbox}
\usepackage[table]{xcolor}
\usepackage{subcaption}
\usepackage{caption}
\usepackage{graphicx,epsfig,sidecap,wasysym}
\usepackage{color,epstopdf}
\newcommand{\media}[1]{\langle #1 \rangle}

\usepackage[normalem]{ulem}
\usepackage[colorlinks,bookmarks=false,citecolor=blue,linkcolor=blue,urlcolor=blue]{hyperref}

\newcommand{\be}{ \begin{equation} }
\newcommand{\ee}{\end{equation}}
\newcommand{\bea}{ \begin{eqnarray} }
\newcommand{\eea}{\end{eqnarray}}

\journal{Annals of Physics Special Issue: Localisation 2020}

\begin{document}

\begin{frontmatter}

\title{Hilbert-space fragmentation, multifractality, and many-body localization}

\author{Francesca Pietracaprina}
\address{School of Physics, Trinity College Dublin, Dublin 2, Ireland}
\author{Nicolas Laflorencie}
\address{Laboratoire de Physique Th\'eorique, Universit\'e de Toulouse, CNRS, UPS, France}

\begin{abstract}
Investigating many-body localization (MBL) using exact
numerical methods is limited by the exponential
growth of the Hilbert space. However, localized
eigenstates display multifractality and  only extend over a vanishing fraction of the Hilbert space.
Here, building on this remarkable property, we develop a simple yet efficient decimation scheme to discard the irrelevant parts of the Hilbert space of the random-field Heisenberg chain. This leads to an  Hilbert space fragmentation in small clusters,
allowing to access larger systems at strong disorder.
The MBL transition is quantitatively predicted, together with a geometrical interpretation of MBL multifractality as a shattering of the Hilbert space.
\end{abstract}

%\begin{keyword}
%Many-body localization, Hilbert space fragmentation, Exact diagonalization, disordered spin chains
%\end{keyword}

\end{frontmatter}

\tableofcontents

\section{Introduction} Many-Body localization (MBL) is one of the most intriguing phenomena of condensed matter physics~\cite{jacquod_emergence_1997,gornyi_interacting_2005,basko_metal-insulator_2006}. While being the natural extension of the celebrated Anderson localization problem to interacting particles, it appears to be conceptually much more difficult to grasp, as compared to its non-interacting counterpart~\cite{evers_anderson_2008}. Nonetheless, during the past decade an enormous amount of theoretical works (for recent reviews, see~\cite{nandkishore_many-body_2015,abanin_recent_2017,alet_many-body_2018,abanin_many-body_2019}) and impressive experimental achievements~\cite{schreiber_observation_2015,ovadia_evidence_2015,de_luca_dynamic_2015,choi_exploring_2016,smith_many-body_2016,kaufman_quantum_2016,roushan_spectroscopic_2017,xu_emulating_2018,wei_exploring_2018,chiaro_growth_2019,lukin_probing_2019} have deeply explored several aspects of MBL physics.

It is now well admitted~\cite{pal_many-body_2010,serbyn_local_2013,huse_phenomenology_2014,luitz_many-body_2015,vosk_theory_2015,potter_universal_2015,imbrie_diagonalization_2016,imbrie_local_2017} that in one dimension, a large class of quantum interacting systems displays a disorder-induced dynamical transition at high energy between two radically   different regimes. At low disorder, high-energy eigenstates are ergodic and thermal in the sense that they obey the eigenstate thermalization hypothesis (ETH)~\cite{deutsch_quantum_1991,srednicki_chaos_1994}, they display high (volume-law) entanglement~\cite{page_average_1993,kjall_many-body_2014,luitz_many-body_2015}, and are fully ergodic in the Hilbert space (HS)~\cite{mace_multifractal_2019,10.21468/SciPostPhysCore.2.2.006}. Conversely, at strong disorder ETH fails and  eigenstates are only area-law entangled~\cite{bauer_area_2013} (a property usually restricted to ground-states~\cite{eisert_area_2010,dupont_many-body_2019}). Moreover, a generic HS multifractality has been observed~\cite{luca_ergodicity_2013,luitz_many-body_2015,tikhonov_many-body_2018,mace_multifractal_2019,solorzano2021multifractality}: more precisely, for a given $\cal N$-dimensional HS, MBL eigenstates only spans a vanishing fraction of it $\sim {\cal N}^{D}$, with $D<1$~\cite{luca_ergodicity_2013,luitz_many-body_2015,mace_multifractal_2019}.

This potentially huge reduction of the support of many-body localized eigenstates, as compared to the full HS, calls for the development of a controlled decimation scheme in order to efficiently discard the irrelevant part of the HS, thus promising a potentially significant computational gain in our description of MBL physics. Furthermore, besides such numerical considerations, a better understanding of the very structure of the HS is conceptually of prime interest. Indeed, as we show in this paper, the MBL phenomenon is rooted in a fragmentation of HS. Some related phenomena, sometimes dubbed HS shattering, have been discussed in other contexts, such as
constrained systems~\cite{PhysRevB.101.174204,PhysRevLett.124.207602,moudgalya2019thermalization,PhysRevB.100.214313,PhysRevResearch.2.023159,herviou2020manybody},  quantum many-body scarred models~\cite{PhysRevLett.124.180604,vanvoorden2020disorder,langlett2021hilbert}, or
dipole-conserving Hamiltonians~\cite{PhysRevX.10.011047,PhysRevB.101.125126}. In the case of quenched disorder, interesting analogies between MBL and HS percolation pictures have also been explored~\cite{roy_exact_2018,roy_percolation_2019,PhysRevB.103.045139,deng2021fragment}. One should also mention that such a Fock space point of view has also motivated  a lot of studies on random graphs and random matrices~\cite{de_luca_anderson_2014, tikhonov_anderson_2016, tikhonov2016fractality, PhysRevB.96.201114, garcia-mata_scaling_2017, biroli2018delocalization, kravtsov2018non, tikhonov2019statistics, garcia-mata_two_2020,PhysRevB.101.134202,PhysRevB.102.014208,PhysRevLett.125.250402,PhysRevResearch.2.043346,detomasi2020rare,Biroli_2021}.

In this work, building on the eigenstate fractality properties, we establis an HS decimation scheme~\cite{aoki_real-space_1980,monthus_many-body_2010} combined with state-of-the-art exact diagonalization (ED) techniques~\cite{luitz_many-body_2015,pietracaprina_shift-invert_2018}. This leads to the following main result:  at strong disorder an MBL Hamiltonian can be practically studied on a vanishing fraction of the original HS, thus allowing an improvement in the accessible system sizes. Surprisingly, one can even quantitatively capture the MBL transition for the random-field Heisenberg chain model~\cite{pal_many-body_2010,luitz_many-body_2015} although this decimation framework is approximate by nature. We further provide a quantitative study of the HS fragmentation by analyzing the disorder-induced development of disconnected HS clusters whose scaling can be directly related to the multifractal dimensions~\cite{mace_multifractal_2019}.

\section{Model} We start with the well-studied random-field Heisenberg spin-$\frac{1}{2}$ ring, described by
\begin{equation}
 {\cal H}=\sum_{i=1}^L \left( S^x_i S^x_{i+1} + S^y_i S^y_{i+1}  + S^z_i S^z_{i+1} + h_i S^z_i\right),
 \label{eq:realspaceH}
\end{equation}
where $h_i$ is drawn from a uniform distribution in $[-h,h]$.
The total magnetization being conserved, we choose the largest subspace of zero magnetization, $\sum_i S^z_i=0$, of total size
\be
\mathcal N=\binom{L}{L/2}\approx \frac{2^L}{\sqrt{{\pi L/2}}}.
\ee
The most recent numerical studies on finite chains predict an MBL transition  in the middle of the spectrum
for $h_c\sim 4$~\cite{luitz_many-body_2015,mace_multifractal_2019,chanda_time_2020,PhysRevResearch.2.032045,PhysRevResearch.2.042033}. Note however that some debates are still active regarding the precise location of this critical point~\cite{devakul_early_2015,doggen_many-body_2018,chanda_time_2020}.

The many-body Hamiltonian Eq.~\eqref{eq:realspaceH} can be mapped onto a tight-binding single particle problem~\cite{welsh_simple_2018,logan_many-body_2019,mace_multifractal_2019} on a high-dimensional lattice built out of its basis states
\begin{equation}
 {\cal H}=\sum_{j=1}^{\mathcal N} \mu_j \ketbra{j}{j} + t \sum_{\langle j,k\rangle} \ketbra{j}{k},
 \label{eq:tightbindingH}
\end{equation}
where the spin configuration basis $\{\ket{j}\}_{j=1,\ldots,\,\cal N}$ is the local projection of  $S^z_i$, e.g. $\ket{\uparrow\downarrow\uparrow\downarrow\cdots}_{z}$ and all its combinations. The first diagonal term in Eq.~\eqref{eq:tightbindingH} is an on-site inhomogeneous potential built from  interactions and random fields
\be
\mu_j=\langle j|\sum_{i}S^z_i S^z_{i+1} + h_i S^z_i|j\rangle,
\ee
while the second term stands for a constant hopping $t=1/2$ between neighboring spin configurations connected by  transverse spin couplings  $S_i^x S_{i+1}^x+S_i^y S_{i+1}^y$.

% structure

In this form, $\cal H$ is nothing but the adjacency matrix of a complex graph whose vertices $j$ are basis states, weighted by $\mu_j$ and connected by edges of constant strength $t=1/2$. The vertex degree, determined by the number of flippable spin pairs $\uparrow\downarrow$ of the corresponding basis state $\ket{j}$, is given on average by $\langle z\rangle =(L+1)/2$ and its distribution approaches a gaussian of variance $\propto L$.
For finite disorder strength $h$, on-site energies $\mu_j$ have a normal distribution of variance $\sigma_\mu^2\approx a h^2 L$~\footnote{We numerically estimate the prefactor $a=0.0824(5)$.}. The fact that the ratio between the mean degree $\media{z}\sim L$ and the effective disorder strength $\sigma_\mu\sim h \sqrt{L}$ diverges in the limit of infinite system size excludes a genuine Anderson localization in this configuration space. Instead, and contrary to random graphs with fixed connectivity~\cite{garcia-mata_scaling_2017,garcia-mata_two_2020}, here a multifractal regime takes over in the MBL phase above $h_c$, where only a subextensive part of the HS is exploited~\cite{luca_ergodicity_2013,luitz_many-body_2015,mace_multifractal_2019}.

%%%%%%%
\begin{figure}[b!]
\centering
 \includegraphics[width=.85\columnwidth]{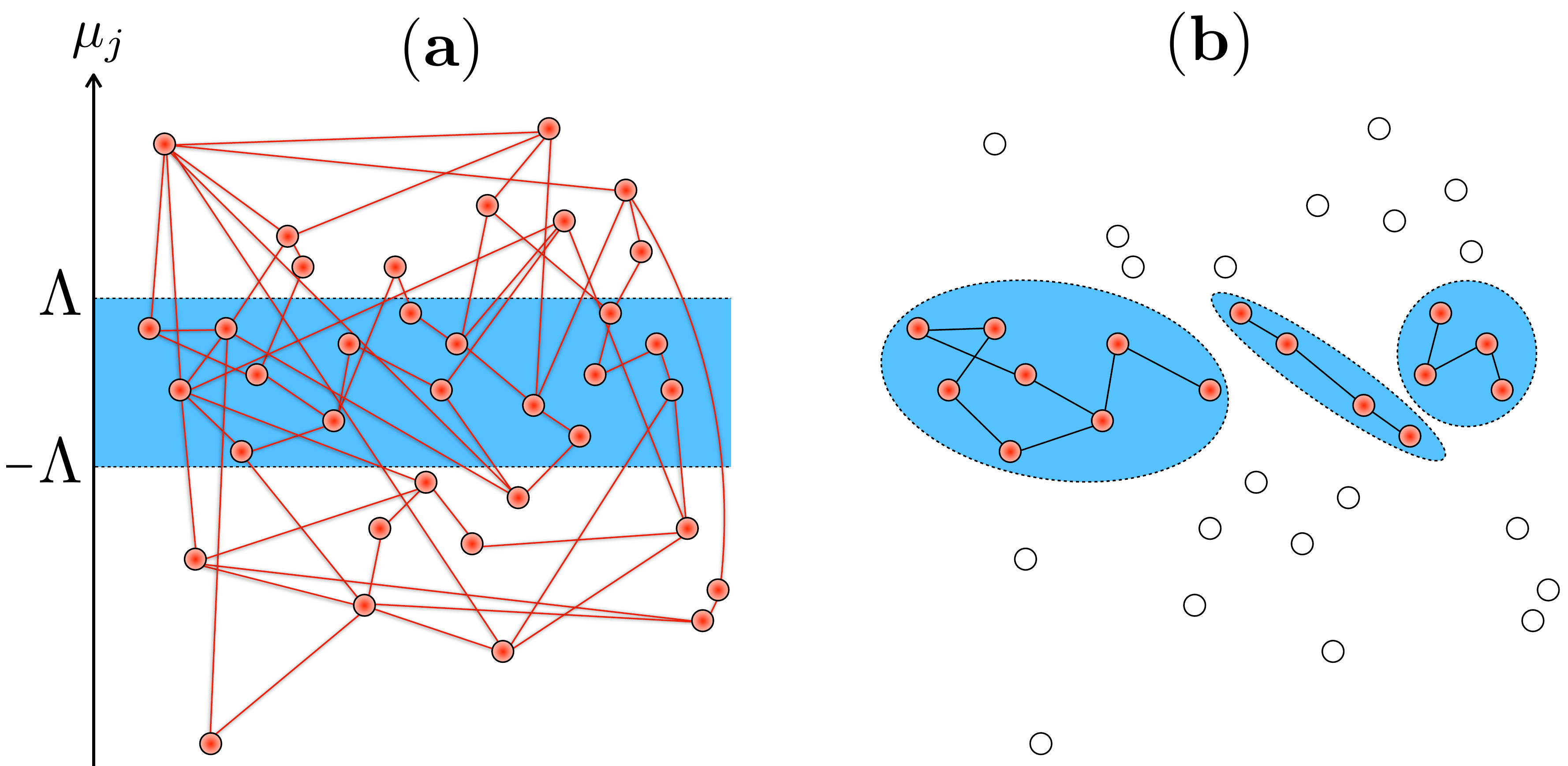}
\caption{Simplified picture for the Hilbert space fragmentation of model Eq.~\eqref{eq:realspaceH} at strong disorder in the MBL regime. (a) Basis states $|j\rangle$ in the tight-binding representation Eq.~\eqref{eq:tightbindingH} are schematized by {vertices} (red circles), connected to neighbors (red {edges}). (b) {Vertices} having on-site energies $|\mu_j|> \Lambda$ are discarded (open circles) and the remaining active Hilbert space clusterizes into smaller pieces (blue bubbles).}
\label{fig:schematic}
\end{figure}
%%%%%%%

\section{Decimation and Hilbert space fragmentation}
\label{sec:fragmentation}
\subsection{Decimation rules}
In order to take advantage, and somehow cure such an ``underutilization'' of HS degrees of freedom, we introduce a decimation scheme, directly acting on the single-particle model Eq.~\eqref{eq:tightbindingH}. This procedure uses the structure of resonances in the tight-binding problem to discard the trivial information that is encoded in the HS in the localized phase. We first note that the locator expansion for the resolvent~\cite{pietracaprina_forward_2016,scardicchio_perturbation_2017}
\begin{equation}
 G_{ab}=\frac{1}{E-\mu_a-\Sigma_{a}^{E}}\sum_{p\,\in\,\text{path}\,(a,b)}\prod_{j\in p}\frac{t}{E-\mu_j-\Sigma_{j}^{E}},
 \label{eq:locator}
\end{equation}
where $E$ is the energy and $\Sigma_j^E$ is the self-energy of site $j$, is convergent in the localized phase. We are interested in the largest terms in this expansion,  which make Eq.~\eqref{eq:locator} diverge, signalling delocalization. If we consider $E=0$ (at the center of the many-body spectrum) and neglect the self-energy, the factors in the product in Eq.~\eqref{eq:locator} are of the type ${t}/{\mu_j}$. Thus, the ones that could contribute to the divergence of the series are those for which $\abs{{t}/{\mu_j}}\gg 1$.

We consider the following decimation procedure: out of all the vertices in the graph generated by $\cal H$, we keep only the ones whose corresponding contribution in Eq.~\eqref{eq:locator} is $\abs{{t}/{\mu_j}}\gg 1$; namely, fixing an $O(1)$ cutoff value $\Lambda>t$, we keep the vertices $j$ whose weight is $\abs{\mu_j}<\Lambda$, see  Fig.~\ref{fig:schematic} (a).
All vertices not satisfying this condition are discarded, as well as all edges connected to them. This removes the sites and paths that are not resonant. The adjacency matrix of the remaining subgraph thus defines the decimated Hamiltonian ${\cal H}_\Lambda$ of size ${\cal N}_\Lambda \times {\cal N}_\Lambda$.  In contrast with Monthus and Garel~\cite{monthus_many-body_2010}, our approach neglects the renormalization of on-site energies, which is justified in the strong disorder limit. Furthermore, we also neglect the renormalization of hoppings, thus restraining the proliferation of new weak bonds. {Incidentally, this enables the appearance of disconnected clusters, which, as it will be shown below, offers insight both from their direct analysis, and from the ED of the restricted Hamiltonian. Moreover we note that,} as recently discussed for the Anderson localization transition and the high-dimensional limit~\cite{tarquini_critical_2017}, the generation of such very small hoppings appears to be irrelevant at strong disorder, a rationale further justified here by the  growing $\propto L$ connectivity and the constant hopping $t=1/2$ of the model.

\subsection{Hilbert space clusterization}
The average fraction of surviving vertices in ${\cal H}_{\Lambda}$ depends on both the disorder strength $h$ and the cutoff $\Lambda$.
Taking advantage of the normal distribution for the on-site energies $\mu_i$, at large enough $L$ it is readily given by
\be
\frac{\langle\mathcal{N}_\Lambda\rangle}{\mathcal{N}}=\erf\left(\frac{\Lambda}{\sigma_\mu\sqrt{2}}\right)\approx b\frac{\Lambda}{h\sqrt{L}},
\ee
with $b\approx 2.78$. From this scaling we immediately envision the potential numerical gain at strong disorder, with an effective HS size $\langle\mathcal{N}_\Lambda\rangle$ reduced by a factor $\propto h\sqrt{L}$, albeit the dominant $2^L$ scaling remains. However, a closer inspection of the decimated HS geometry reveals a much more interesting effect: while a giant percolating cluster exists at low disorder, ${\cal H}_\Lambda$  gets fragmented in disconnected components at high disorder (schematized in Fig.~\ref{fig:schematic}),
a related phenomenon also observed in Refs.~\cite{roy_exact_2018,roy_percolation_2019} {where a classical percolation transition in configuration space signals} the MBL transition.
Conversely here our clusterization mechanism is a non-universal process which depends on the cutoff value $\Lambda$~\footnote{Note that our approach differs from Refs.~\cite{roy_exact_2018,roy_percolation_2019} where bonds with $|\mu_i-\mu_j|>t$ are discarded.}.

\subsubsection{Largest component}
\begin{figure*}[ht!]
\includegraphics[width=\columnwidth]{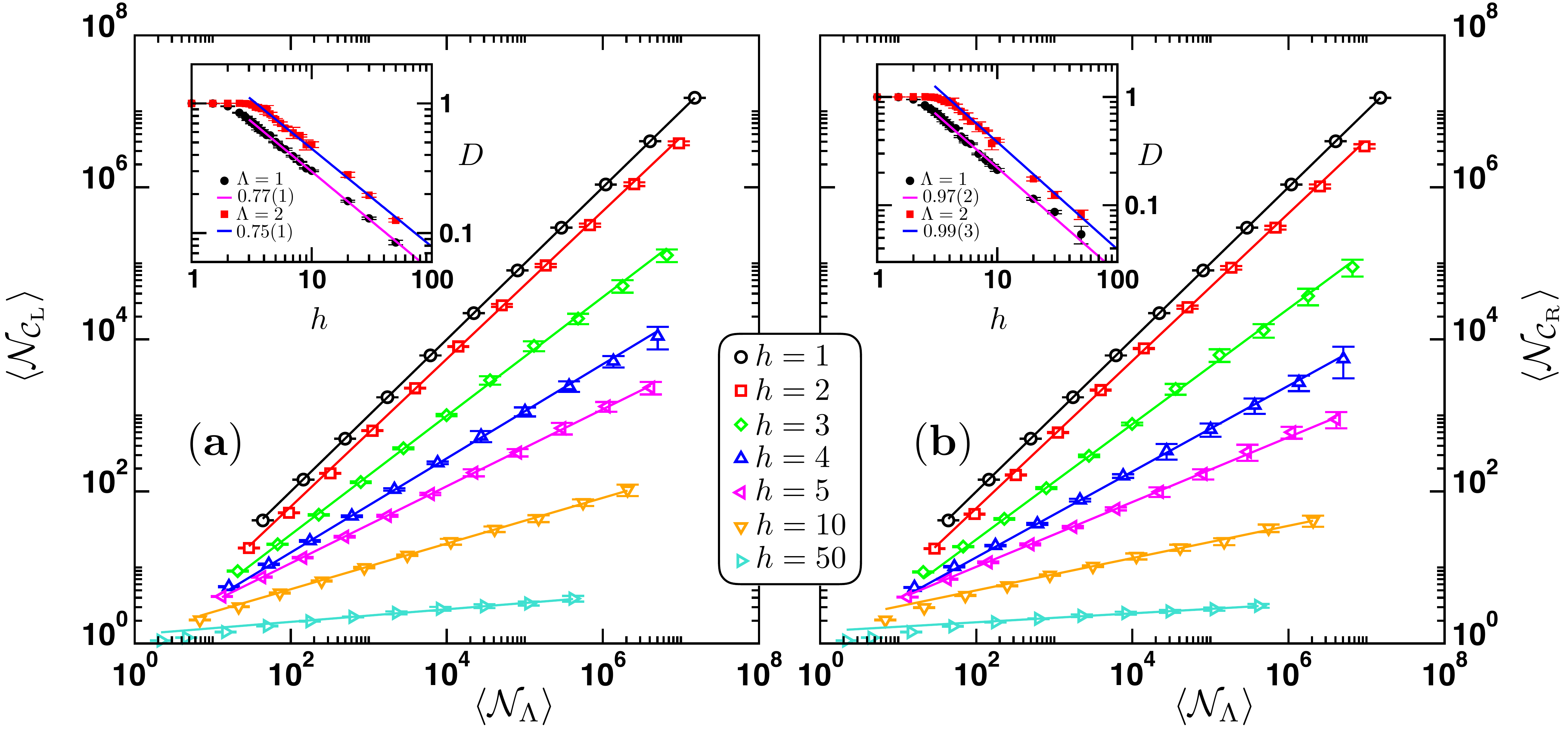}
\caption{Power-law scaling of the average cluster sizes as a function of the average cut HS size $\langle{\cal N}_\Lambda\rangle$, for the largest (a) and random (b) clusters, shown for a few representative values of the disorder strength $h$. Lines are fits to the form $\langle {\cal{N}}_\Lambda\rangle^{D}$, where $D$ is the fractal dimension, plotted against $h$ in the insets for the two cases, and $\Lambda=1,\,2$. A power-law decay is observed at large enough $h$, $D(h)\sim h^{-\beta}$ with $\beta_{\rm L}\approx 0.75$ for $\mathcal{C}_{\rm L}$ and $\beta_{\rm R}\approx 1$ for $\mathcal{C}_{\rm R}$.\label{fig:clusters}}
\end{figure*}

{We find a power-law distribution for the cluster sizes at strong disorder.} We first focus on the largest cluster $\mathcal{C}_{\rm L}$ which is identified after the enumeration of  all components in ${\cal H}_\Lambda$, for various spin chain sizes, up to $L=28$, corresponding to an original HS of size ${\cal N}\sim 4\times 10^7$~\footnote{The enumeration is performed using the breadth-first search algorithm, whose complexity scales as $L\times {\cal{N}}_{\Lambda}$.}, {and averaged over many disorder realizations (at least $1000$ and up to $2\cdot10^4$}). Its disorder-averaged size $\langle{\cal N}_{{\cal C}_{\rm L}}\rangle$ is shown in Fig.~\ref{fig:clusters} (a) for various values of disorder strength and a cutoff  $\Lambda=1$. We observe a clear power-law scaling with $\langle{\cal{N}}_\Lambda\rangle$:
\be
\langle {\cal N_C}\rangle \propto \langle {\cal{N}}_\Lambda\rangle ^{D_{\Lambda,h}},
\label{eq:power}
\ee
where the exponent $D_{\Lambda,h}\le 1$ has a non-trivial $\Lambda$ and $h$-dependence, shown in the inset of Fig.~\ref{fig:clusters} (a). Indeed, the pseudo-critical disorder $h^*$ for which $D<1$, signalling the fragmentation, clearly depends on $\Lambda$, thus contrasting with the universal percolation mechanism found in Refs.~\cite{roy_exact_2018,roy_percolation_2019}. Nevertheless, at strong disorder one finds a power-law decay
\be
D_{\Lambda,h}\sim h^{-\beta},
\label{eq:D}
\ee
with $\beta_{\rm L}\approx 0.75$ for $\Lambda=1,\,2$.

%%%%%%%%%%%%%%%%%%%%%%%%%%%%%%%%%%%%%%%
%%%%%%%%%%%%%%%%%%%%%%%%%%%%%%%%%%%%%%%

\subsubsection{Typical clusters}
The largest cluster is clearly not a typical one, rather capturing rare events.  We therefore apply another scheme to analyze the clusterization.
For each disordered sample we pick up a `random' cluster, $\mathcal{C}_{\rm R}$, chosen in the following way: we consider the set of vertices with highest degree in ${\cal H}_\Lambda$ and select one vertex randomly; we then consider the component to which the selected vertex belongs to.
This cluster can be constructed very efficiently (linear in the cluster size $\cal{N}_{\mathcal{C}_{\rm R}}$), which strongly reduces the high computational cost of the cluster enumeration we had to apply before to identify the largest (untypical) component. Moreover,  random clusters will be more representative and typical, as compared to the largest one.

The average size $\langle{\cal N}_{\mathcal{C}_{\rm R}}\rangle$ is shown in Fig.~\ref{fig:clusters} (b), where one also observes a power-law  behavior vs. $\langle {\cal N}_\Lambda\rangle$, Eq.~\eqref{eq:power}. Here the exponent $D_{\Lambda,h}$ shows a similar non-trivial  $\Lambda$ and $h$-dependence, as displayed in the inset of Fig.~\ref{fig:clusters} (b) with a decay at strong disorder Eq.~\eqref{eq:D} occuring with a different exponent $\beta_{\rm R}\approx 1$. Nevertheless, the departure from $D=1$ also depends on $\Lambda$, showing no evidence of {a possible correspondence between a percolation transition in the HS and} the MBL transition.

At this stage, it is useful to make a link with recent ED results obtained for the multifractal scalings in the MBL regime~\cite{mace_multifractal_2019}, where it was found that MBL eigenstates are supported by only a sub-extensive portion of the configuration space $\propto {\cal N}^{D_{\rm MBL}}$ with $D_{\rm MBL}\sim 1/h$ at strong disorder. Here, our decimation scheme reaches similar conclusions, with
a strong disorder decay of the fractal dimension $D_{\Lambda,h}$ of ${\cal C}_{\rm R}$ following Eq.~\eqref{eq:D} with an exponent  $\beta_{\rm R}\approx 1$, in agreement with $D_{\rm MBL}$. This result gives a striking geometrical interpretation of the MBL eigenstate multifractality observed in Ref~\cite{mace_multifractal_2019}, and the sub-extensive HS portion is understood as an HS fragmentation.

\section{{Exact diagonalization of random clusters}}
To go beyond geometrical considerations, we aim at exploring in more details the microscopic behavior of eigenstates on such random clusters. Taking advantage of the reduced HS size at high disorder, larger system sizes (as compared to the state-of-the-art~\cite{pietracaprina_shift-invert_2018}) can be accessed numerically by applying the shift-invert ED method~\cite{luitz_many-body_2015,pietracaprina_shift-invert_2018} on the corresponding matrices ${\cal H}_{\Lambda}$ and ${\cal H}_{{\cal C}_{\rm R}}$.

\subsection{Participation entropy}
In order to explore the wave functions properties, we compute the participation entropy, defined for an eigenstate ${\ket\Psi}$ in the spin basis $\{\ket{j}\}$ by
\begin{equation}
{S_{\rm P}}=-{\sum_jp_j\ln p_j},\label{eq:SP}
\end{equation}
where $p_j=|\langle j|\Psi\rangle|^2$ is the probability of occupation for each state $\ket j$. ED data, obtained for a decimation cutoff  $\Lambda=1$, are averaged over many samples ($10^4$ realizations for $L\leq 18$, $10^3$ for $18<L\leq26$, $500$ for $L=28$) and a few eigenstates in the middle of the spectrum. {We note that states within the clusters, especially when the latter are small (that is, at very high disorder and relatively small size), can be far from to the middle of the spectrum of the full Hamiltonian. We select only clusters that contain states within $5\%$ of the center of the spectrum.}

\subsection{Numerical results}
\begin{figure}[t!]
\centering
\includegraphics[width=.875\linewidth]{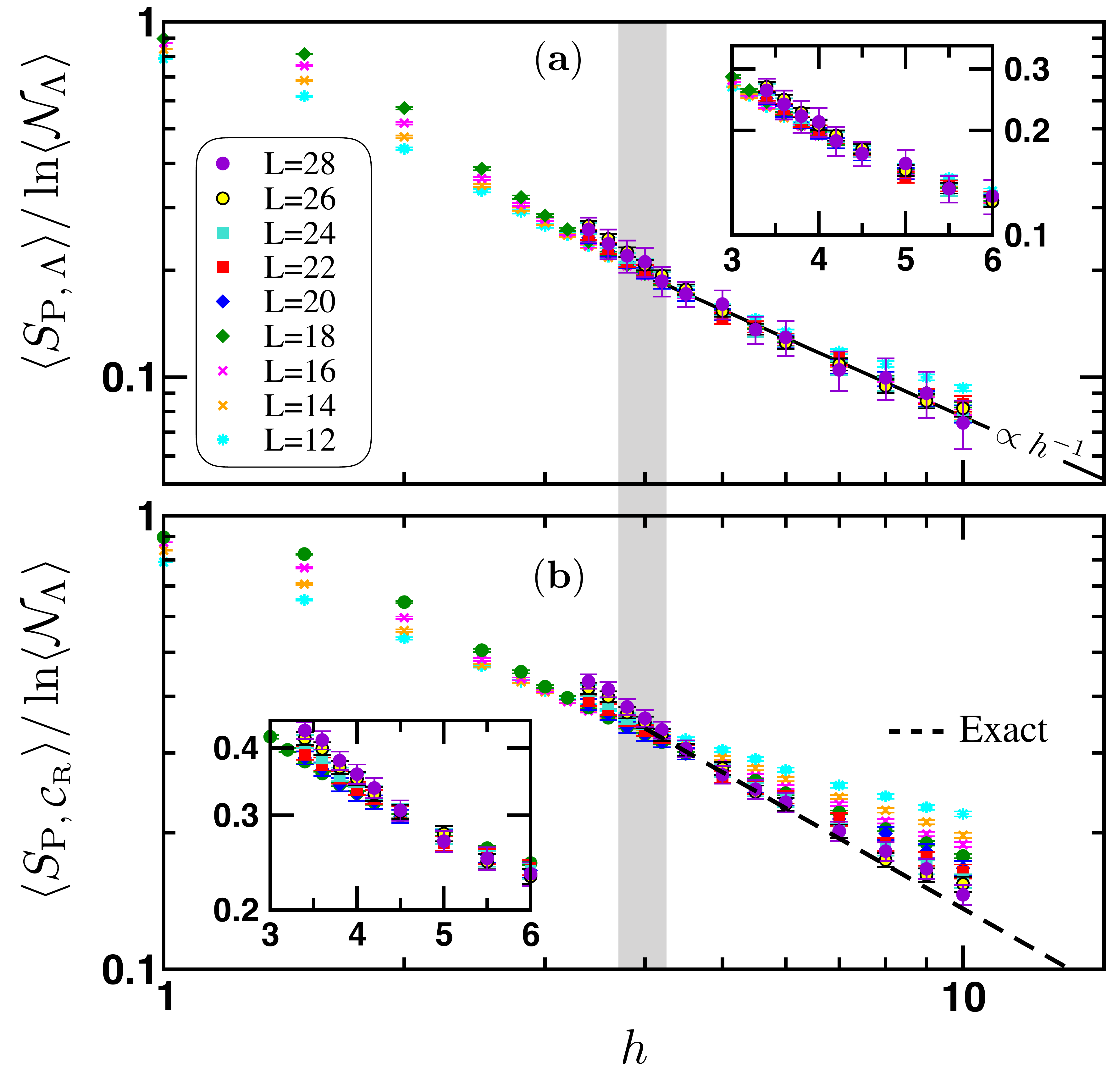}
\caption{Participation entropy Eq.~\eqref{eq:SP}, rescaled by the decimated HS size $\ln\langle {\cal N}_{\Lambda}\rangle$ with $\Lambda=1$, is plotted as a function of disorder strength $h$. Data, averaged over disorder and a few eigenstates in the middle of the spectrum, are shown for both (a) ${\cal H}_{\Lambda}$ and (b) ${\cal{H}}_{{\cal{C}}_{\rm R}}$, for various system lengths $L=12,\ldots,\,28$ as indicated on the graph. {The black line in the top panel is $\propto h^{-1}$. We additionally show as a dotted line the exact results from ED of the original Hamiltonian $H$ (from Ref.~\cite{mace_multifractal_2019}).} The MBL transition is signalled by a crossing of the curves, which occurs in the grey region for $h_c\sim 4$. Insets are zooms over the crossing regions.}
\label{fig:Sp}
\end{figure}

Following recent results ~\cite{luca_ergodicity_2013,luitz_many-body_2015,mace_multifractal_2019}, we expect the disorder-average participation entropy
$\langle S_{\rm P}\rangle$ to grow as $D\ln \mathcal{N}_\Lambda$, with $D=1$ in the delocalized phase, and $D\sim 1/h$ deep in the MBL regime~\cite{mace_multifractal_2019}.
In Fig. \ref{fig:Sp} we show $\langle S_{\rm P}\rangle$, rescaled by $\ln \langle{\mathcal N}_\Lambda\rangle$, as a function of disorder for two cases: (a) the full decimated Hamiltonian ${\cal H}_{\Lambda}$, and (b) for random clusters ${\cal{H}}_{{\cal{C}}_{\rm R}}$. The results reproduce the expected behavior~\cite{mace_multifractal_2019} in the MBL regime with the correct $h^{-1}$ dependence, but also near the transition and in the delocalized phase. In addition, the MBL transition can be estimated almost quantitatively by the crossing of the data close to $h_c\sim 4$, in quite good agreement with the best ED estimates~\cite{luitz_many-body_2015,mace_multifractal_2019,PhysRevResearch.2.032045,PhysRevResearch.2.042033}. It is worth noting however, that a slight overestimation of the transition point {as well as a possible weak $\Lambda$ dependence can be} expected due to the enhancement of the delocalization effects by neglecting the self-energies in Eq.~\eqref{eq:locator}. {Qualitatively,} similar results are obtained for both ${\cal H}_{\Lambda}$ and ${\cal{H}}_{{\cal{C}}_{\rm R}}$. {Notably, ${\cal{H}}_{{\cal{C}}_{\rm R}}$ gives rise to values that are extremely similar to the ED of the original Hamiltonian $H$, as well as more pronounced and more realistic finite size corrections.}

We find that MBL eigenstates properties are well captured by our decimation scheme. Indeed, removing a vertex $j$ in the graph is equivalent to impose a vanishing wavefunction amplitude on the corresponding configuration basis state $\ket j$. While not exact, this turns out to reproduce fairly faithfully the microscopic structure of eigenstates in the MBL regime, as well as close to the critical point.
On the other hand, the properties of the energy spectrum, e.g. the level statistics, are very sensitive to the HS fragmentation which induces a block structure in ${\cal H}_{\Lambda}$, leading to an emergent integrability. Therefore, one would observe the appearance of a Poisson distribution for the energy gaps at a non-universal, cutoff-dependent, pseudo-critical disorder $h^*(\Lambda)$, as observed in the insets of Fig.~\ref{fig:clusters} for $D$.

\section{Discussions and conclusions} In this work, we have proposed a simple yet efficient decimation scheme to address the MBL problem, working directly in the spin configuration basis of the random-field Heisenberg chain. Naturally designed for strong disorder, this approach is able to give quantitative results in the MBL regime, and can even capture the transition of the random-field Heisenberg chain at high energy.

{We highlight once more the potential numerical gain that is accessible with this decimation scheme followed by ED. Although the exponential scaling of the Hilbert space with the real-space system size can not be escaped, we have shown a way to reduce the scaling from $\sim2^L$ to a much more favorable $\sim2^{DL}$, where $D\approx 0.1$. Given current numerical ED limits, and assuming a careful basis construction, we show in Table~\ref{tab:sizes} the physical sizes for which it is possible to study the Heisenberg chain model Eq.~\eqref{eq:realspaceH} in the ${\cal{C}}_{\rm R}$ cluster and with a cutoff parameter $\Lambda=1$.}

\begin{table}
\centering
\begin{tabular}{l|cccccc}
  \diaghead{\theadfont Lhspace+}%
{h}{L}& \thead{16} & \thead{20} & \thead{24} & \thead{28} & \thead{32} & \thead{40} \\
  \hline\\
  2 & \cellcolor{green!25} $2\cdot 10^3$ & \cellcolor{green!25} $3\cdot 10^4$ & \cellcolor{green!25} $3 \cdot 10^5$ & \cellcolor{yellow!25} $3\cdot 10^6$ & \cellcolor{red!25} $\lesssim 6 \cdot 10^7$ & \cellcolor{red!25} $\lesssim 9 \cdot 10^9$\\
  3 & \cellcolor{green!25} $3\cdot 10^2$ & \cellcolor{green!25} $2\cdot 10^3$ & \cellcolor{green!25} $1\cdot 10^4$ & \cellcolor{yellow!25} $9\cdot 10^4$ & \cellcolor{yellow!25} $\lesssim 10^6$& \cellcolor{red!25} $\lesssim 5 \cdot 10^7$\\
  4 & \cellcolor{green!25} $80$  & \cellcolor{green!25} $3\cdot 10^2$ & \cellcolor{green!25} $1 \cdot 10^3$ & \cellcolor{green!25} $6\cdot 10^3$ & \cellcolor{green!25} $\lesssim 5 \cdot 10^4$ & \cellcolor{yellow!25} $\lesssim 10^6$ \\
  5 & \cellcolor{green!25} $30$ & \cellcolor{green!25} $10^2$ & \cellcolor{green!25} $3 \cdot 10^2$ & \cellcolor{green!25} $9\cdot 10^2$ & \cellcolor{green!25} $\lesssim 9 \cdot 10^3$ & \cellcolor{green!25} $\lesssim 10^5$ \\
  10 & \cellcolor{green!25} $8$ & \cellcolor{green!25} $14$ & \cellcolor{green!25} $20$ & \cellcolor{green!25} $40$ & \cellcolor{green!25} $\lesssim 2 \cdot 10^2$ & \cellcolor{green!25} $\lesssim 8 \cdot 10^2$
 \end{tabular}
\caption{Average size of the matrix ${\cal{H}}_{{\cal{C}}_{\rm R}}$ for various spin chain length and disorder values. The green colored cells correspond to matrix sizes that are easily accessible with full ED, the yellow ones are accessible with shift-invert ED, while the red cells are not trackatable.}
\label{tab:sizes}
\end{table}

A simple picture of HS fragmentation emerges for the MBL regime, thus providing a straightforward geometrical interpretation to the eigenstate multifractality, albeit with no clear evidences for a percolation scenario~\cite{roy_exact_2018,roy_percolation_2019}. The possible links between this HS fragmentation and the recently discussed spin freezing observed at strong disorder~\cite{PhysRevResearch.2.042033} remain to be explored. From a methodological point of view, we have illustrated the feasibility and the efficiency of the method up to $L=28$ (i.e. a gain of one order of magnitude for the HS size as compared to standard ED~\cite{pietracaprina_shift-invert_2018}), but this approach can be further improved to reach larger systems at strong disorder. This opens an avenue to investigate the existence MBL in two dimensions~\cite{choi_exploring_2016,bordia_probing_2017,thomson_time_2018,wahl_signatures_2019,de_tomasi_efficiently_2019,theveniaut_many-body_2019}.

\section{Acknowledgement}
We are  grateful to Fabien Alet, Gabriel Lemari\'e, Nicolas Mac\'e, C\'ecile Monthus for comments and interesting discussions.
This work benefited from the support of the project THERMOLOC ANR-16-CE30-0023-02 of the French National Research Agency (ANR). We acknowledge CALMIP (grants 2017-P0677 and 2018-P0677) and GENCI (grant x2018050225) for HPC resources.
\newpage
%\bibliography{mbl}

\end{document}